# Simple all-optical method for in situ detection of ultralow amounts of ammonia

Yuanchao Liu[a], Tristan Asset[a], Yechuan Chen[a], Eamonn Murphy[a], Eric O. Potma[b], Ivana Matanovic[c], Dmitry A. Fishman[b*], Plamen Atanassov[a]

[a]*Department of Chemical & Biomolecular Engineering, National Fuel Cell Research Center (NFCRC), University of California, Irvine, California 92697, USA*

[b]*Department of Chemistry, University of California, Irvine, California 92697, USA*

[c]*Department of Chemical and Biological Engineering, University of New Mexico, Albuquerque, New Mexico 87131, USA*

## Abstract

As a key precursor for nitrogenous compounds and fertilizer, ammonia affects our lives in numerous ways. Rapid and sensitive detection of ammonia is essential, both in environmental monitoring and in process control for industrial production. Here we report a novel and nonperturbative method that allows rapid detection of ammonia at detection levels of only a few thousand molecules, based on the non-contact, all-optical detection of surface-enhanced Raman signals. We show that this simple and affordable approach enables ammonia probing at selected regions of interest with high spatial resolution, making *in situ* and *operando* observations possible.

* Corresponding author: dmitryf@uci.edu

## Introduction

Ammonia is key to many chemical technologies that drive world economies[1, 2], being the major source for fertilizers for agriculture/food production[3] and directly used in multitude of industrial processes[4]. Ammonia analyses are critical for water quality control[5], exhaust gas sensing[6, 7], monitoring and process control in industries of new chemicals, plastics and pharmaceutical. It is also critical in detection of combustibles[8], detonation and propulsion processes. Moreover, it is at heart of our quest to find alternative routes of nitrogen fixation mechanisms for large-scale ammonia generation, that are more efficient, energy saving and carbon dioxide footprint free[9]. In all of these processes, rapid and sensitive detection of ammonia is critical, thus driving a need for reliable ammonia sensing technologies. Current detection concepts are based on various "wet chemistry" processes and devices - electrochemical sensor devices[10-12], metal oxides[13-20], catalytic polymer[21-28], inkjet printing films[29], polymer hybrid framework[30]. Spectroscopic methods for ammonia detection include nuclear magnetic resonance (NMR)[9, 31] and optical spectroscopy in almost every part of electro-magnetic spectrum, ranging from the ultraviolet[32] to the far IR branch[33-38].

Among the full arsenal of ammonia detection methods[12], spectroscopic approaches are of particular interest as they provide direct information on chemical content and unambiguously identify the presence of the target molecular compound even at low concentrations. Traditionally, nuclear magnetic resonance spectroscopy (NMR) has been the method of choice for sensitive ammonia detection[9, 31]. NMR provides chemical selectivity and high sensitivity down to 3 μM of the target and it is rightfully among the most trusted approaches. However, complex instrumentation and large required sampling volumes (>0.5 mL) constitute serious limitations. More importantly, prompt to contamination, the NMR approach relies on using isotope labeling, which renders even routine experiments very expensive and limits applications mainly to fundamental research in laboratory environments.

At present, the most popular and widely adopted optical methods are based on colorimetric approaches and their experimental derivatives[39]. A popular solution employs the Berthelot reaction between ammonia, chlorine and phenolic compounds, resulting in blue coloration of an indophenol dye that can be easily detected by conventional spectrometers[35]. This approach allows ammonia detection with an impressive sensitivity, down to several tens of ppb (parts per billion) [35, 36, 40, 41]. While considered as the "gold standard", the blue indophenol method requires time-consuming sample handling. An aliquot of >1 mL has to be collected from the chemical reactor and mixed with toxic, sample-altering reagents, followed by tens of minutes of aging prior to spectroscopic analysis.

In this context, the availability of a detection method for ammonia and its derivatives that is robust, affordable, and universally applicable would be highly desirable. We envision that *simplicity, speed, sensitivity, affordability and reliability* are five crucial ingredients to advance ammonia sensing and making it more accessible for both laboratory and field use. Existing spectroscopic methods do not allow tracking of small changes in the concentration of ammonia and intermediate species on-site, not to mention *in situ* or *operando* observation in situations where the reaction mechanisms are not known. Most spectroscopic techniques and sensor approaches are directly or indirectly vulnerable to external laboratory or field environment, jeopardizing *reliability*. While NMR and colorimetric methods excel in detection limits for overall concentration, it is important to remember that both approaches rely on the need for relatively large sensing volumes. For instance, strong signals are observed for a 1 ml volume at <1 ppm (parts per million) concentrations, which contains $10^{15}$-$10^{17}$ molecules of ammonia. For fundamental research, this means even greater number of molecules has to be produced at reactor heart to fulfill volume and sensitivity requirements for further analysis. This obstacle is elegantly overcome using surface-sensitive IR absorption spectroscopy, which is based on the principle of total attenuated reflection (ATR-FTIR)[42, 43]. This approach uniquely probes molecules produced right on the surface of electrode. It is a relatively fast and non-perturbative technique, requiring only tens of seconds for spectral acquisition. At the same time, the fixed experimental geometry, the necessity to isolate the IR light from the laboratory environment and the need for additional chemicals limit a broader implementation of the method.

In this work, we present a novel approach to ammonia detection, enabled through the use of surface-enhanced Raman scattering (SERS). This is a simple, fast yet sensitive method that provides chemical selectivity to ammonia and does not rely on complex sensor design, specific beam path geometry or chemical modifications of the solution. The method allows observation of amount as little as $10^4$-$10^5$ molecules that can be detected locally at the region of interest, which paves the way to ultrasensitive *operando* electrochemical experiments. The following corresponds to sub 1 ppm concentrations to be measured in just under 1s.

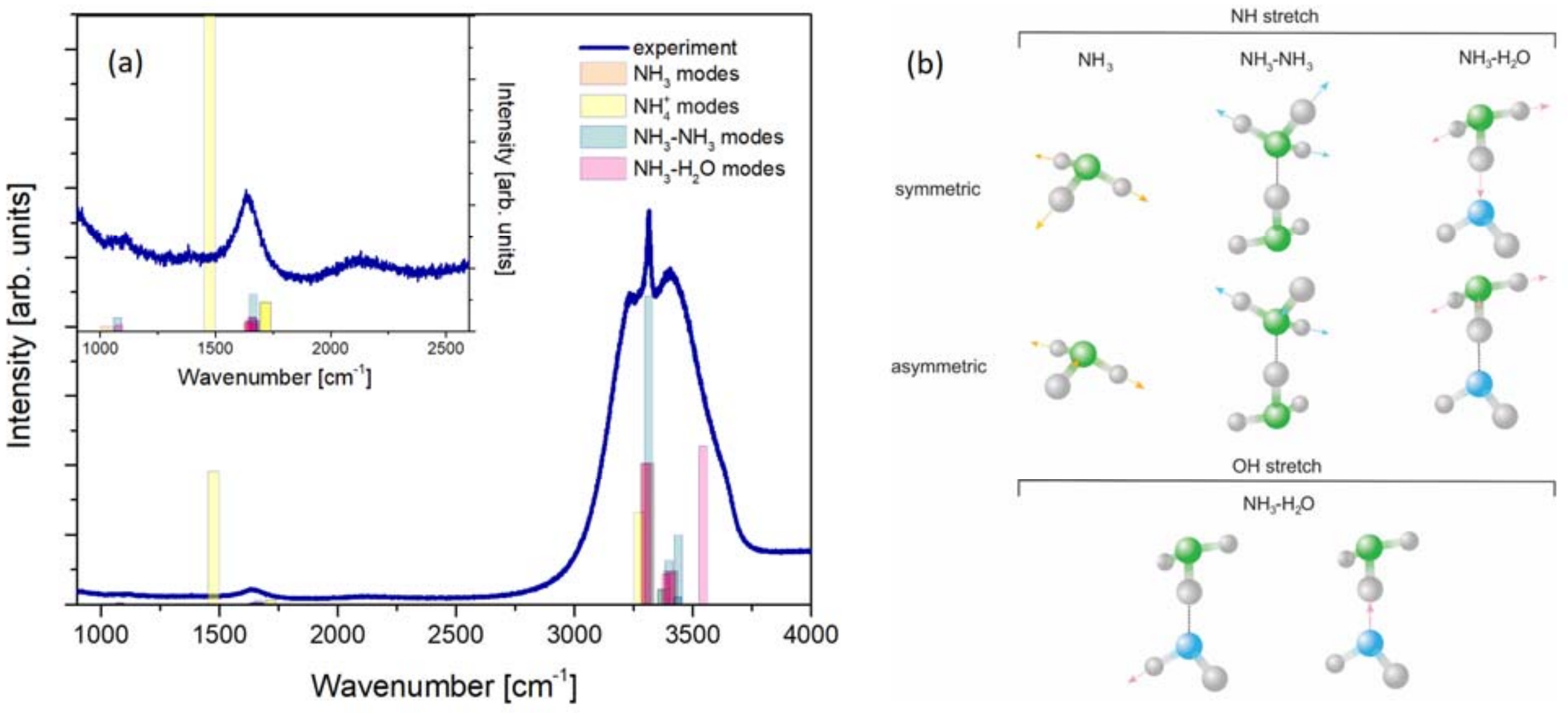

**Figure 1.** (a) Spontaneous Raman spectrum of bulk ammonia solution (0.3 wt%, 30000 ppm). Column bars indicate positions of theoretically predicted bending, stretching and rocking modes of ammonia related formations. (b) main ammonia related vibrations presented in stretching modes spectral region.

## Results

Figure 1a shows a representative spontaneous Raman spectrum of a concentrated ammonia solution spanning the molecular fingerprint region and the stretching modes of N-H, O-H and C-H related complexes. To understand the signatures of this spectrum, we performed a normal mode frequency analysis for ammonia and its aqueous formations via density functional theory (DFT) calculations, visualized in Figure 1b. These formations are addressed thoroughly in previous works and are most widely accepted contributors to vibrational spectrum of ammonia aqueous solution[37, 38, 44-51] - ammonia ($NH_3$), ammonium ion ($NH_4^+$), ammonia dimer ($NH_3$-$NH_3$), ammonia-water complex ($NH_3$-$H_2O$) and solvated ammonia complex (not shown in Figure 1). Positions and relative intensities of the observed lines and their corresponding theoretical predictions are in excellent agreement. For more details on the DFT calculations and a table of the mode frequencies, see *Methods* and *Supplementary Information*.

Though not very bright, several lines from monomer and dimer complexes are clearly visible in the fingerprint spectral region (Figure 1a, inset). Theory shows that, in contrast to the ammonia molecule, the ammonium ion should have the strongest presence in this frequency range through a 1476 $cm^{-1}$ contribution of a N-H deformation mode. Yet, the experimental absence of this line is confirmed through the evaluation of the $NH_3$ / $NH_4^+$ mole ratio by pH measurements (see *Methods*). It should be noted that even 100 ppb ammonia dissolved in neutral pH solution consists of 95% ammonia molecules compared to ammonium ions. This agrees well with spectroscopy data, where only a broad line comprised of H-N-H bending of $NH_3$, $NH_3$-$NH_3$ and $NH_3$-$H_2O$ complexes is observed.

| Formation | Spectral position | Intensity | Assignment |
|---|---|---|---|
| $NH_3$ | 3409.0 $cm^{-1}$ | 36.8 | NH asymmetric stretching |
| | 3309.8 $cm^{-1}$ | 155.8 | NH symmetric stretching |
| $NH_4$+ | 3380.0 $cm^{-1}$ | 16.8 | NH asymmetric stretching |
| | 3276.8 $cm^{-1}$ | 101.0 | NH symmetric stretching |
| $NH_3$-$NH_3$ | 3441.0 $cm^{-1}$ | 76.2 | NH asymmetric stretching |
| | 3438.0 $cm^{-1}$ | 7.8 | NH asymmetric stretching |
| | 3402.3 $cm^{-1}$ | 48.2 | NH asymmetric stretching |
| | 3369.5 $cm^{-1}$ | 15.6 | NH asymmetric stretching |
| | 3315.4 $cm^{-1}$ | 341.6 | NH symmetric stretching |
| $NH_3$-$H_2O$ | 3545.6 $cm^{-1}$ | 175.0 | OH stretching (H-bonded) |
| | 3410.9 $cm^{-1}$ | 35.3 | NH asymmetric stretching |
| | 3389.2 $cm^{-1}$ | 33.0 | NH asymmetric stretching |
| | 3305.0 $cm^{-1}$ | 155.31 | NH symmetric stretching |

**Table 1.** Main ammonia related vibrations presented in stretching modes spectral region. For full mode list, please, see Supplementary Table ST1.

In contrast to the weak vibrational features in the fingerprint range, the stretching modes reveal clear signatures of ammonia's presence with intensities that are two orders of magnitude higher than fingerprint modes, in full agreement with DFT predictions. The origin of the peaks in this spectral region and their positions has been thoroughly studied [16, 37, 38, 46, 48] with the main peaks outlined in Table 1. From the perspective of relative intensities, the highest Raman signal corresponds to the N-H symmetric stretching mode that clearly stands out against the broad and complex O-H stretching band of water. According to DFT calculations, the asymmetric N-H stretching has 4 to 5 times smaller Raman intensity, but strongly contributes to the broad profile of the overlapping water signals. Ujike *et al.* obtained the normal mode frequencies of the ammonia molecule, ammonia-ammonia dimer and ammonia-water dimer by means of the GF matrix method[38]. In this method, the constants of the force matrix were determined from a direct fitting of the experimental data. In contrast, our approach utilizes force derivate matrices determined from first principles and as such relies on basic physical parameters of the compounds involved. For the ammonia monomer molecule, we obtained excellent agreement between this approach and previous works[38, 52]. However, for the complex dimer formations there are significant discrepancies. While the lowest energies for ammonia and ammonia-water dimers have been determined at 3217.4 $cm^{-1}$ and 3044.1 $cm^{-1}$, respectively[38], first principles anharmonic frequency calculations for dimer modes of $NH_3$-$NH_3$ and $NH_3$-$H_2O$ do not differ by more than 10 $cm^{-1}$. Moreover, our theoretical study did not find any ammonia normal modes at the lower energy side from the main peak at 3260 $cm^{-1}$. The broad profile on which the main peak rests can be assigned to a distribution of additional contributions from solvated ammonia species, *i.e.* molecules that are engaged through multiple hydrogen bonds. At the high energy side of the band, our predictions match previous estimates for ammonia-water complexes with non-bonded and bonded O-H stretches at 3989 $cm^{-1}$ and 3721 $cm^{-1}$ [48] (not shown in Figure 1b, see Supplementary Table ST1).

The strong peaks from the symmetric N-H stretch at ~3300 $cm^{-1}$ provide an opportunity to employ these spectroscopic signatures for tracing ammonia molecules or related reaction pre-cursors. However, due to the presence of the substantial water background and the weak response in the fingerprint range, Raman spectroscopy has long been considered unsuitable for ammonia detection. Indeed, regular Raman scattering is blind to solutions with concentrations below 1000 ppm (Figure SF1). To heighten and discriminate molecular signatures from the strong aqueous signals we employed surface-enhanced Raman scattering (SERS) – a well-known approach to detect analytes down to single molecule limit[53, 54]. The broadly accepted electromagnetic theory for enhancement relies on excitation of localized surface plasmons (LSP), which provide strongly localized fields that can enhance the optical response of molecules in their vicinity. Surprisingly, SERS has not been specifically utilized for ammonia detection, though some basic principles of the effect in relation to ammonia has been demonstrated in 1984 in work by Sanchez *et al.*[55] Further studies on detection of ammonium nitride revealed sensitive detection limits in fingerprint branch [56], even though all the observed and discussed lines are not related to $NH_3$, but rather $NO_3^-$. Recently the possibility of ammonia detection with excitation on the slope of plasmonic resonance has been discussed [57, 58]. In this work, an elegant implementation of SERS detection for live reaction monitoring was discussed with acquisition time constants of ~10 seconds. While the authors focused mainly on the fingerprint region and sidebands from probable ammonia-water dimers, no stretching peak of the $NH_3$ monomer has been observed in *operando* experiments.

*Ammonia detection with commercial SERS substrates*

For our first initial demonstration of ultrasensitive ammonia detection via SERS we used a conventional confocal microscopy setup and commercially available SERS substrates (SERStrate, Silmeco, Denmark). The substrates consist of leaning silicon nanopillars covered with silver[59]. The structure forms a rather dense and homogeneous distribution of metal covered clusters of 50 nm – 100 nm with corresponding plasmonic resonances that closely match the excitation wavelength of 532 nm. The aqueous ammonia solution was confined between the SERS substrate and the microscope slide, with the Raman signal collected in the epi-geometry by a water immersion objective and coupled into a CCD-based spectrometer (Figure 2a, see Methods for details). The setup was used in two regimes – confocal and wide-field approach (Figure 2b and 2c). The confocal collection geometry allows probing of fine local ammonia concentrations at the region of interest. For the given experimental conditions and the numerical aperture of the collection objective, this volume corresponds to only ~1.2 $\mu m^3$ or 1.2 fL of solution being probed. To enable excitation within a wider region of interest, the beam was focused at the back aperture of the excitation/collection objective and the imaging aperture/slit was effectively removed from the image plane. The latter allowed for an increased focal volume by almost an order of magnitude, as well as a more effective use of the detector area without significantly losing spectral resolution. Hence, this wide-field imaging mode greatly improves detection sensitivity (Figure SF2).

A typical SERS spectrum of an aqueous ammonia solution is shown in Figure 3, acquired using confocal geometry and at only 200 ppm concentration (Figure 2b). Strong ammonia SERS signals clearly overcome the water background compared to the spontaneous Raman signal of the bulk solution. The bright line around ~3260 $cm^{-1}$ indicates high sensitivity to the symmetric N-H stretching mode with an improvement of the detection limit by a thousand-fold relative to regular Raman spectroscopy. The spectral shift (~50 $cm^{-1}$) and line broadening are anticipated due to the molecule-surface interaction, with the main line of the $NH_3$ monomer changed from 14 $cm^{-1}$ (for spontaneous Raman) to 30 $cm^{-1}$ (for SERS). Overall, the spectral shape corroborates theoretical predictions and can be closely reproduced with a simple Gaussian broadening of the $NH_3$ monomer and $NH_3$-$H_2O$ dimer modes (see Table ST2). Note the rather strong presence of additional spectral structures around 2930 $cm^{-1}$ that were not visible in spontaneous Raman measurements of bulk solution (Figure SF2a), but are found at certain locations on the substrate through SERS (Figure SF3). Further experiments confirmed that these signals do not scale similarly with ammonia concentration and N-H related bands. These lines were also occasionally observed in SERS of pure Millipore water. Additional experiments and simple line shape analysis suggest that these signals can be assigned to the strong C-H stretching modes of occasional organic contamination on the plasmonic structure (Figure SF4). These characteristic modes in the C-H stretching spectral range are well known, and can be confidently used for identifying organic species in bio-related environments.[60, 61].

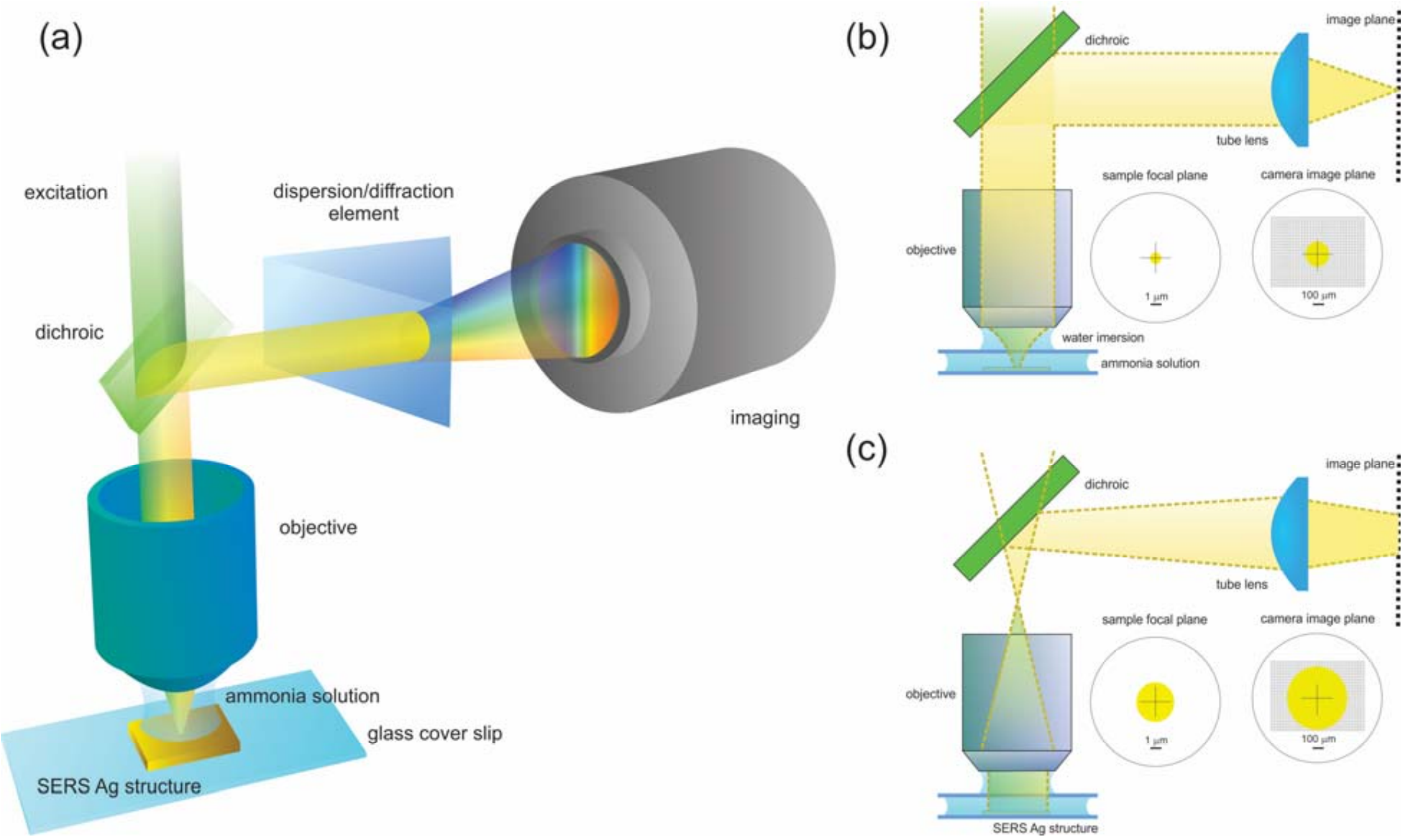

**Figure 2.** Concept of experimental approach. (a) Epi detection of Raman signal, (b) confocal detection for high spatial resolution, (c) wide-field detection approach for large volume probing.

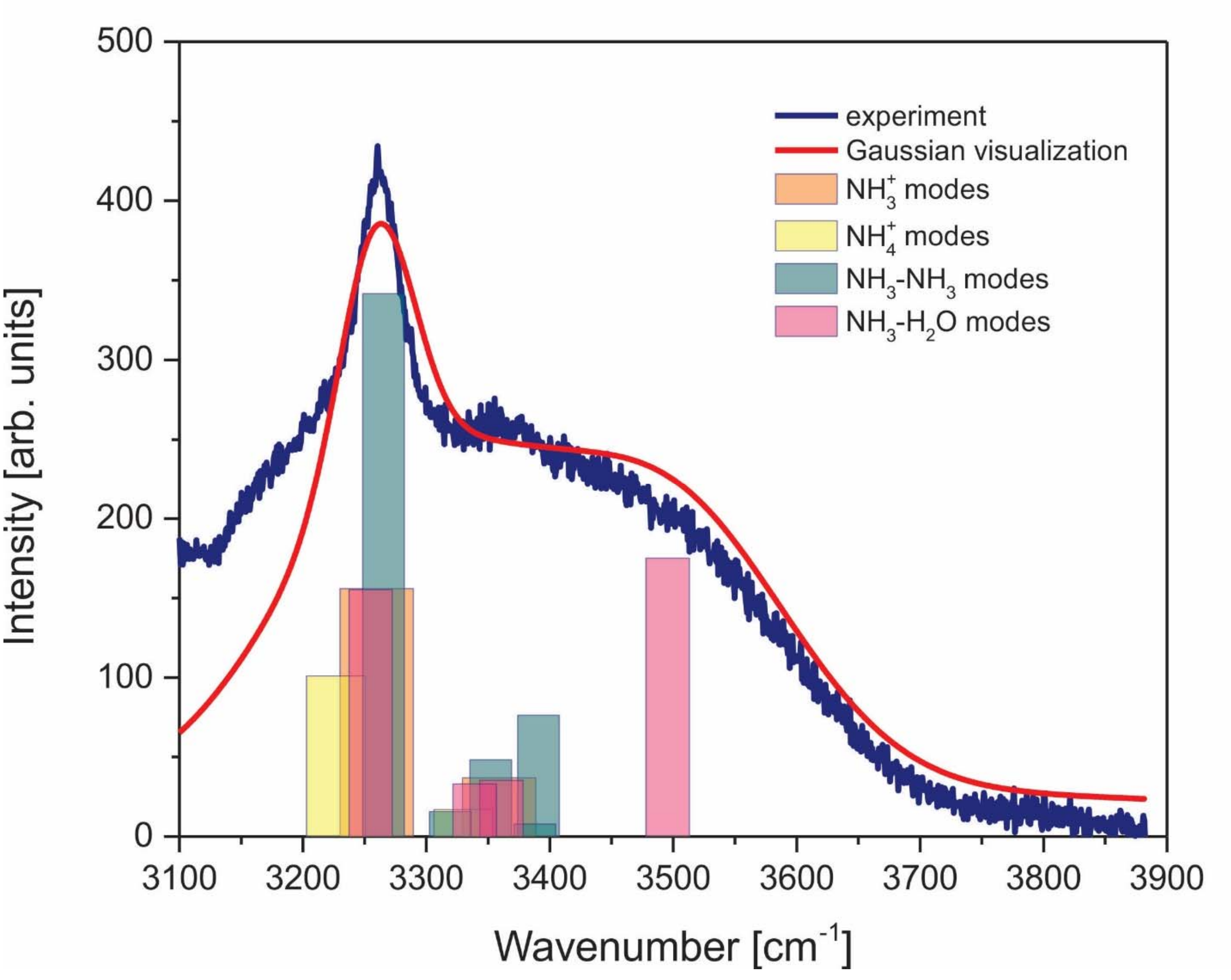

**Figure 3.** Calculated modes for ammonia related structure overlaid with SERS spectrum of 200 ppm ammonia solution. Red curve is visualization through using Gaussian line shapes at predicted theoretical positions for $NH_3$ and $NH_3$-$H_2O$ only. Exposure time 0.3 s.

The SERS signal dependence on the ammonia concentration gives rather important insights into the actual chemical content of the solution. As observed in Figure 4a, for higher concentration solutions the ammonia-related bands dominate the bulk water background with broad and rich spectral features in the stretching mode region. With a decrease of concentration, the spectrum evolves into a steady plateau – the broad O-H spectrum of bulk water molecules. A detailed analysis of the integrated intensity for the bright spectral feature around 3260 $cm^{-1}$ reveals clear linear behavior as a function of ammonia concentration (Figure 4b). The following means that signals in this range are dependent on complexes that contain a single $NH_3$ molecule, *i.e.* the $NH_3$ monomer and the $NH_3$-$H_2O$ dimer, as any signal response from $NH_3$-$NH_3$ dimers is anticipated to be quadratic. We stress that all observations are fully reversible when concentration levels are changed on the same SERS substrate, as shown in Figure SF5. This points to the absence of a strong chemical interaction with the metal surface, and thus allows reversible use of the sensor over a large dynamic range of the ammonia concentration.

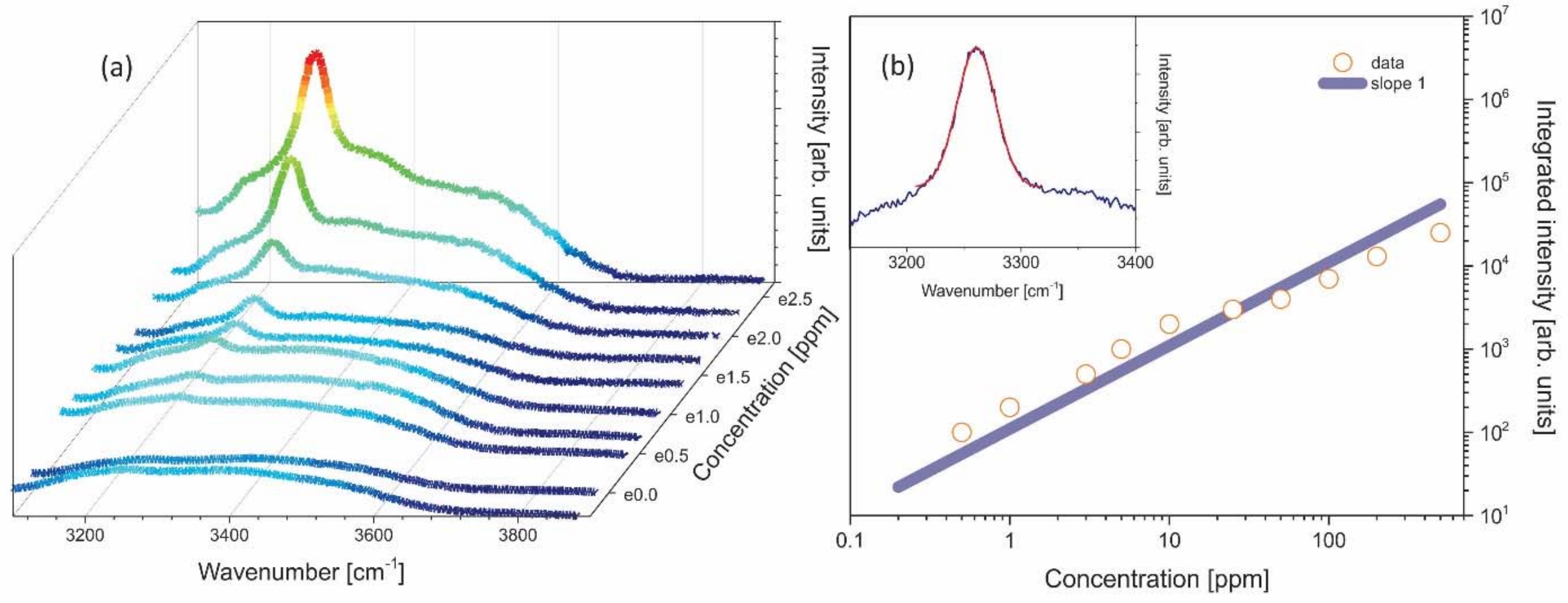

**Figure 4.** Concentration dependence studies. (a) Spectral evolution as a function of ammonia concentration. (b) Integrated intensity of 3260 cm-1 spectral line. Blue straight line represents linear function with slope 1. Inset: Gaussian fit of ammonia related line for 200 ppm solution.

Figure 4 and Figure 5 illustrate the detection sensitivity of the approach. We note that all spectra have been acquired without any signal processing, modulation or electronic filtering. Acquisition times did not exceed 2 s to reveal the true and unperturbed potential of the method.

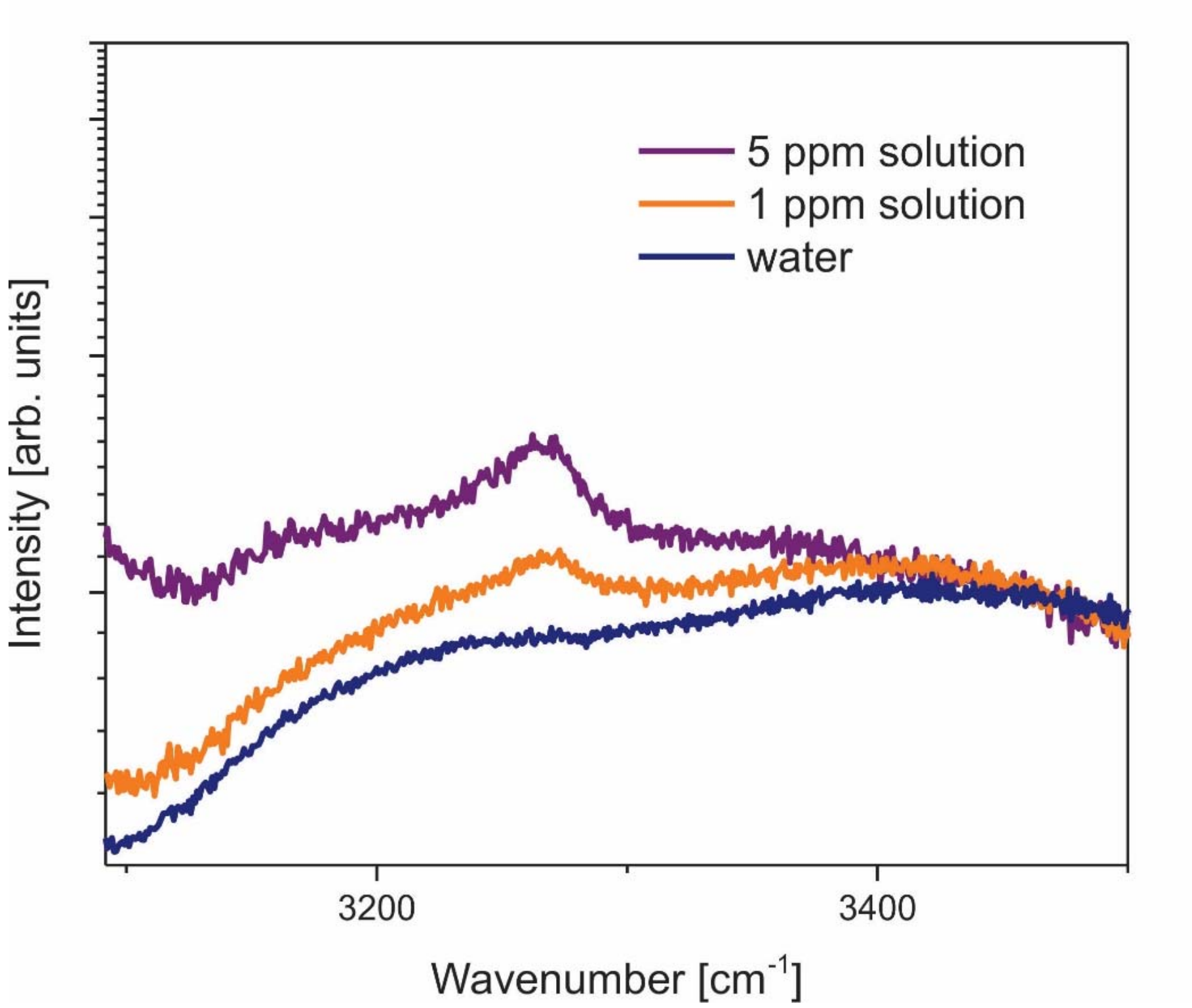

**Figure 5.** Fast detection of low amount of ammonia using commercial Ag-coated Si nanopillars substrate. Exposure time 1s.

For the given plasmonic structure, collection efficiencies and experimental geometries, 1 ppm concentrations of ammonia can be clearly observed for measurement times of 1 s or less. For longer acquisition times of 1 to 2 seconds the detection limit of few hundreds of ppb is easily attainable (Figure 5). Besides the sensitivity in overall ammonia concentration, another useful metric for sensitivity is the actual number of molecules probed. For the confocal geometry experiments with a 1.2 fL probe volume, a 1 ppm concentration corresponds to only $10^4$-$10^5$ molecules – the amount necessary for spectroscopic observation (Table ST3). Though this molecular detection limit is already significantly smaller compared to competing methods, these numbers are still conservative estimates. The estimation relies on the total amount of molecules in the confocal volume with Rayleigh range of about ~1.5 μm. In reality, only those molecules that are in near-field proximity (<1 nm) to the metal structure are contributing to the signal. Thus, the actual number of sampled molecules may still be a few orders of magnitude smaller. In the wide-field illumination approach, the surface area efficiently increases just under ten times, thus probing a greater number of molecules contained within the detection area.

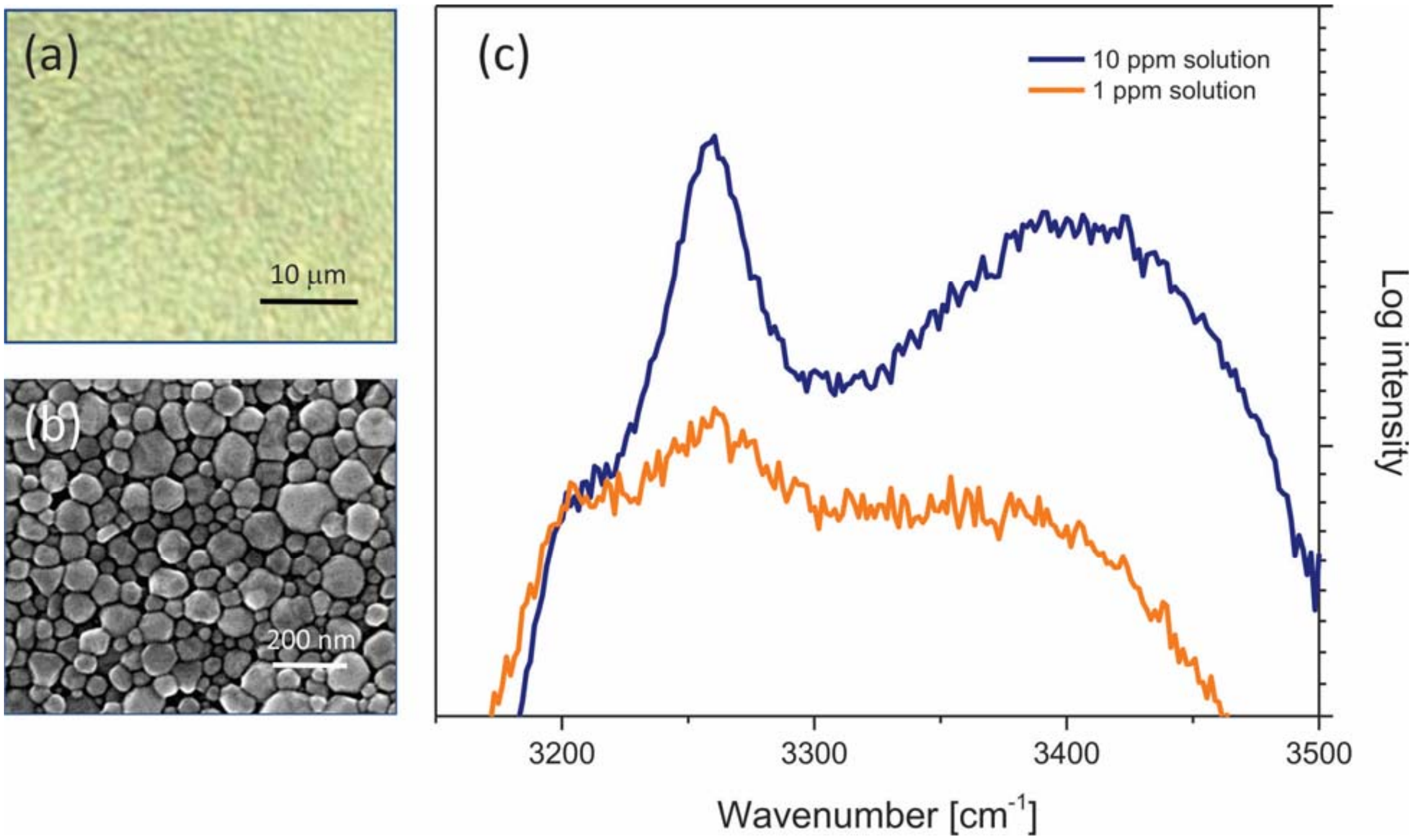

**Figure 6.** Fast detection of low amount of ammonia using drop casted Ag ink on a glass slide. (a) Optical image of dried Ag ink, (b) SEM image of dried Ag ink, (c) SERS signal on custom made SERS substrate. Exposure time 1 s.

*Ammonia detection on Ag ink*

We have performed further experiments based on a similar detection concept, but that employs less sophisticated plasmonic structures to explore more affordable sensor alternatives. For this purpose, we used a commercially available Ag ink (Sigma Aldrich), based on metal nanoparticles. This material is widely used for printable conductors, antibacterial material filters, thin film electronics, *etc*. This particular ink is based on silver particles ($d_{90}$ = 115 nm, $d_{50}$ =70 nm) with plasmonic resonances that match our previous experiments. When drop casted and dried on mica glass slide, the colloidal suspension forms a homogeneous mirror-like film. Optical microscopy and SEM images of the film are shown in Figure 6a and Figure 6b, respectively. Despite its simplicity, the ink-based SERS substrate permits clear observations of signals with a sub-1ppm detection limit for ammonia. Although the enhancement, spatial distribution and reproducibility are not on par with commercial SERS substrates, the approach can be improved by adopting better controlled and cleaner printing procedures.

## Discussion

The SERS method discussed in this work offers a new approach for detecting ammonia and related complexes with a chemical sensitivity that is competitive with that of other spectroscopic techniques, while at the same time being straightforward and affordable in its implementation.

While for the *low limit of detection* the method is already on par with other spectroscopic techniques, in terms of actual number of molecules sampled for producing detectable signals it performs >10 orders of magnitude better than most other techniques. The approach is capable to enter the "single molecule" detection regime, as concentrations below 100 ppb correspond to only a hundred molecules within the near-field interaction distance of the sampled region of interest. However, such sensitivity does not come at the expense of long acquisition time. The acquisition *speed* easily allows live observations – sub-second detection of only a few thousands of molecules at the region of interest makes *in situ* monitoring of chemical reaction possible.

As an optical-based approach, it allows distant, nonperturbative (non-consuming) probing, thus offering clean and *reliable* experiments, insensitive to contamination from spurious ammonia of the surrounding environment. Moreover, molecules do not appear to strongly bind to the surface, making observations *reversible* as the molecular content is changed.

The practical implementation of the concept introduced in this work can be significantly improved, in particular in terms of speed and sensitivity. The SERS signal enhancement factors rely on the interplay between spectral position of the plasmonic resonance, excitation source wavelength and the intrinsic spectral response of the molecule. Thus, fine-tuning of the metal particle size and the structure arrangement may result in a significant improvement of detection limits. Whereas commercially available SERS substrates are excellent initial choice, they require complex production and are currently too expensive for up-scaling. We demonstrate that simple deposition of colloidal silver ink on the substrate of choice already gives comparable results. Hence, the method does not require special facilities and procedures to produce *affordable* SERS sensors.

Lastly, the *simplicity* of the method is of key significance to broad implementation. With flexible beam positioning, measurements can be performed at any region of interest within a given environment or chemical cell. Through point scanning or a wide-field imaging mode, the method allows sampling of the spatial distribution of ammonia content. This flexibility has the potential to provide important insights in underlying reactions mechanisms, which is especially relevant for solid-liquid interfaces. In addition, in our experiments with confocal signal collection we have used no more than 1.5 mW CW laser light, light dose that are easily attained with compact and inexpensive laser diode modules. Furthermore, the method can be adapted for use with lower numerical aperture lenses, whereby the lower collection efficiency is offset by a larger illumination area. By probing a greater number of molecules in larger spot sizes, long focusing approach may prove to be advantageous for sensing needs that do not require spatial confinement and resolution.

As underlined in this work, the Raman spectrum in the N-H stretching region is rich in $NH_3$ related signals. Detection of this broadband spectral region, as opposed to the narrow $NH_3$ monomer line alone, will result in drastic improvements of detection limits. The large Stokes shift of this spectral window makes it possible to use simple colored glass Schott filters, rather than expensive sharp-edge Raman dichroic mirrors. Finally, conventional single-pixel Si detectors can be used. Though cheap and affordable, such detectors can match standard charged-coupled devices, when using photo-multipliers and avalanche mode detectors can even boost the sensitivity. All these potential simplifications will allow the design of a simple and robust device to be used for field research or in-line production.

Ammonia detection method through SERS converges *simplicity* and *affordability* with *speed* and *sensitivity*. It has potential to be greatly popularized and help to overcome several roadblocks that has been slowing fundamental and applied chemical research in these areas for many decades.

## Materials and methods

**Solution preparation and pH measurement.** Samples of ammonia solutions were prepared by diluting 56 wt% ammonium hydroxide ($NH_3 \cdot H_2O$) solution (VWR) with *Millipore* water. As results, concentrations of 0.1 ppm, 0.5 ppm, 1 ppm, 3 ppm, 5 ppm, 10 ppm, 20 ppm, 50 ppm, 100 ppm, 200 ppm, 500 ppm, 1000 ppm and 0.3 wt% (30, 000 ppm) ammonia solutions were made, based on the weight of $NH_3$ monomer and $H_2O$ molecules.

In order to estimate the ratio of ammonia molecule ($NH_3$) and ammonium ions ($NH_4^+$), the pH values of ammonia solutions were measured. Specifically, the pH of pure Millipore water was measured at ~6.6 due to slight dissolving of ambient carbon dioxide. The pH values of 0.1 ppm, 10 ppm and 500 ppm solution were 10.7, 10.2 and 11.1 respectively. The $NH_3$/ $NH_4^+$ ratio was simply estimated by the acidity of ammonium (pKa = 9.25), according to:

$$\log_{10}(NH_3/NH_4^+) = pH - pKa$$

As results, the ammonia molecule ($NH_3$) content of all dilute ammonia samples were above 95 %, as compared to ammonium ions ($NH_4^+$).

**Density Function Theory calculation.** The calculation of vibrational modes and Raman intensities were performed using Gaussian09 quantum chemistry program[62] at the M06-2X/aug-cc-pVTZ level of theory[52]. The structures of the ammonia molecule, ammonium ion, ammonia-ammonia dimer and ammonia-water dimer were first optimized using tight convergence criteria and ultrafine integration grid after which the harmonic frequencies were calculated. To account for the anharmonic effects, which influence the frequencies of the stretching modes the most, anharmonic frequencies of the stretching modes were calculated using numerical differentiation along corresponding vibrational modes as implemented in the Gaussian09 software. To study the change in the Raman spectra of the ammonia molecule and ammonia-water dimer due to the solvation, we also tested a model of fully solvated ammonia. The model was built by solvating ammonia-water dimmer with additional 12 water molecules after which the system was optimized. The number of water molecules was chosen to create a cluster in which ammonia-water dimmer would be fully surrounded by water molecules.

**Plasmonic structures.** Commercial SERS substrates SERStrate (Silmeco, Denmark) has been used. Substrates are based on leaned Si nanopillars covered with Ag resulting in resonance around 550 nm. For custom made silver nanoparticle SERS substrate commercially available Ag ink (Sigma Aldrich) has been used. The ink comprises silver nanoparticles of about 100 nm size and is pre-characterized by the vendor using optical and electron microscopy approaches.

**Raman microscopy.** Raman experiments has been performed on custom modified Renishaw InVia Raman microscope. 532 nm SSDP laser was used as excitation source. All excitation and collection of Rama signal were done using 60x, 1.2NA water immersion objective (Olympus). For wide-field imaging beam was focused on input aperture of objective and slit opened to 100 μm size, corresponding to increase of spectral resolution to 9 $cm^{-1}$. For confocal experiments 1.5 mW at 532 nm has been used at the sample. For wide-field images 30 mW has been used to compensate change of the energy flux.

**Acknowledgments**

This work was supported in part by a subcontract from DOE-EERE Advanced Manufacturing Office award to Sandia National Laboratories (AOP 34920). D.A.F. would like to thank Prof. Vartkess A. Apkarian for inspiring discussions. D.A.F. acknowledges NSF grant CHE-0960179. I. M. wishes to thank the UNM Center for Advanced Research Computing for computational resources. Authors are thankful to Dr. Andrea Perego, National Fuel Cell Research Center (NFCRC), University of California Irvine for the SEM images obtained at Irvine Materials Research Institute (IMRI).

**Conflict of interests**

All authors declare no competing interests.

**Contributions**

D.A.F. and P.A. conceived the idea and supervised the study. Y.L. prepared ammonia solutions, Y.L., T.A. and Y.C. directed general logistics of chemical experiments. I.M. performed DFT calculations, D.A.F. designed experimental concepts, conducted optical experiments, analyzed the data and wrote initial manuscript. All authors took part in thorough discussions of data and contributed to its understanding. All authors contributed to manuscript editing.

## References

1. Haber, F. Thermodynamik technischer Gasreaktionen. (Salzwasser Verlag, Paderborn; 1905).
2. Bosch, C. (1908).
3. Ampuero, S. & Bosset, J.O. The electronic nose applied to dairy products: a review. *Sensors and Actuators B: Chemical* **94**, 1-12 (2003).
4. Chen, J.G. et al. Beyond fossil fuel–driven nitrogen transformations. *Science* **360**, eaar6611 (2018).
5. Association, A.P.H., Association, A.W.W., Federation, W.P.C. & Federation, W.E. Standard Methods for the Examination of Water and Wastewater. (American Public Health Association., 1995).
6. Docquier, N. & Candel, S. Combustion control and sensors: a review. *Progress in Energy and Combustion Science* **28**, 107-150 (2002).
7. Riegel, J., Neumann, H. & Wiedenmann, H.M. Exhaust gas sensors for automotive emission control. *Solid State Ionics* **152-153**, 783-800 (2002).
8. Kohl, D. Function and applications of gas sensors. *Journal of Physics D: Applied Physics* **34**, R125-R149 (2001).
9. Andersen, S.Z. et al. A rigorous electrochemical ammonia synthesis protocol with quantitative isotope measurements. *Nature* **570**, 504-508 (2019).
10. Chang, S.C., Stetter, J.R. & Cha, C.S. Amperometric gas sensors. *Talanta* **40**, 461-477 (1993).
11. Lundström, I., Svensson, C., Spetz, A., Sundgren, H. & Winquist, F. From hydrogen sensors to olfactory images - twenty years with catalytic field-effect devices. *Sensors and Actuators: B. Chemical* **13**, 16-23 (1993).
12. Timmer, B., Olthuis, W. & Berg, A.v.d. Ammonia sensors and their applications—a review. *Sensors and Actuators B: Chemical* **107**, 666-677 (2005).
13. Clifford, P.K. & Tuma, D.T. Characteristics of semiconductor gas sensors I. Steady state gas response. *Sensors and Actuators* **3**, 233-254 (1982).
14. Hübner, H.P. & Drost, S. Tin oxide gas sensors: An analytical comparison of gas-sensitive and non-gas-sensitive thin films. *Sensors and Actuators B: Chemical* **4**, 463-466 (1991).
15. Imawan, C., Solzbacher, F., Steffes, H. & Obermeier, E. Gas-sensing characteristics of modified-MoO3 thin films using Ti-overlayers for NH3 gas sensors. *Sensors and Actuators B: Chemical* **64**, 193-197 (2000).

16. Srivastava, R.K., Lal, P., Dwivedi, R. & Srivastava, S.K. Sensing mechanism in tin oxide-based thick-film gas sensors. *Sensors and Actuators B: Chemical* **21**, 213-218 (1994).
17. Wang, X., Miura, N. & Yamazoe, N. Study of WO3-based sensing materials for NH3 and NO detection. *Sensors and Actuators B: Chemical* **66**, 74-76 (2000).
18. Xu, C., Miura, N., Ishida, Y., Matsuda, K. & Yamazoe, N. Selective detection of NH3 over NO in combustion exhausts by using Au and MoO3 doubly promoted WO3 element. *Sensors and Actuators B: Chemical* **65**, 163-165 (2000).
19. Yamazoe, N. New approaches for improving semiconductor gas sensors. *Sensors and Actuators B: Chemical* **5**, 7-19 (1991).
20. Zakrzewska, K. Mixed oxides as gas sensors. *Thin Solid Films* **391**, 229-238 (2001).
21. Cai, Q.Y., Jain, M.K. & Grimes, C.A. A wireless, remote query ammonia sensor. *Sensors and Actuators B: Chemical* **77**, 614-619 (2001).
22. Chabukswar, V.V., Pethkar, S. & Athawale, A.A. Acrylic acid doped polyaniline as an ammonia sensor. *Sensors and Actuators B: Chemical* **77**, 657-663 (2001).
23. Heiduschka, P., Preschel, M., Rösch, M. & Göpel, W. Regeneration of an electropolymerised polypyrrole layer for the amperometric detection of ammonia. *Biosensors and Bioelectronics* **12**, 1227-1231 (1997).
24. Kukla, A.L., Shirshov, Y.M. & Piletsky, S.A. Ammonia sensors based on sensitive polyaniline films. *Sensors and Actuators B: Chemical* **37**, 135-140 (1996).
25. Lähdesmäki, I., Kubiak, W.W., Lewenstam, A. & Ivaska, A. Interferences in a polypyrrole-based amperometric ammonia sensor. *Talanta* **52**, 269-275 (2000).
26. Lähdesmäki, I., Lewenstam, A. & Ivaska, A. A polypyrrole-based amperometric ammonia sensor. *Talanta* **43**, 125-134 (1996).
27. Nicolas-Debarnot, D. & Poncin-Epaillard, F. Polyaniline as a new sensitive layer for gas sensors. *Analytica Chimica Acta* **475**, 1-15 (2003).
28. Palmqvist, E., Berggren Kriz, C., Svanberg, K., Khayyami, M. & Kriz, D. DC-resistometric urea sensitive device utilizing a conducting polymer film for the gas-phase detection of ammonia. *Biosensors and Bioelectronics* **10**, 283-287 (1995).
29. Lv, D. et al. Enhanced flexible room temperature ammonia sensor based on PEDOT: PSS thin film with FeCl3 additives prepared by inkjet printing. *Sensors and Actuators B: Chemical* **298**, 126890 (2019).

30. Ahmadi Tabr, F., Salehiravesh, F., Adelnia, H., Gavgani, J.N. & Mahyari, M. High sensitivity ammonia detection using metal nanoparticles decorated on graphene macroporous frameworks/polyaniline hybrid. *Talanta* **197**, 457-464 (2019).
31. Nielander, A.C. et al. A Versatile Method for Ammonia Detection in a Range of Relevant Electrolytes via Direct Nuclear Magnetic Resonance Techniques. *ACS Catalysis* **9**, 5797-5802 (2019).
32. Mount, G.H. et al. Measurement of atmospheric ammonia at a dairy using differential optical absorption spectroscopy in the mid-ultraviolet. *Atmospheric Environment* **36**, 1799-1810 (2002).
33. Giovannozzi, A.M. et al. An infrared spectroscopy method to detect ammonia in gastric juice. *Analytical and Bioanalytical Chemistry* **407**, 8423-8431 (2015).
34. Max, J.-J. & Chapados, C. Aqueous ammonia and ammonium chloride hydrates: Principal infrared spectra. *Journal of Molecular Structure* **1046**, 124-135 (2013).
35. Patton, C.J. & Crouch, S.R. Spectrophotometric and kinetics investigation of the Berthelot reaction for the determination of ammonia. *Analytical Chemistry* **49**, 464-469 (1977).
36. Tzollas, N.M., Zachariadis, G.A., Anthemidis, A.N. & Stratis, J.A. A new approach to indophenol blue method for determination of ammonium in geothermal waters with high mineral content. *International Journal of Environmental Analytical Chemistry* **90**, 115-126 (2010).
37. Li, Y. & Keppler, H. Nitrogen speciation in mantle and crustal fluids. *Geochimica et Cosmochimica Acta* **129**, 13-32 (2014).
38. Ujike, T. & Tominaga, Y. Raman spectral analysis of liquid ammonia and aqueous solution of ammonia. *Journal of Raman Spectroscopy* **33**, 485-493 (2002).
39. Searle, P.L. THE BERTHELOT OR INDOPHENOL REACTION AND ITS USE IN THE ANALYTICAL-CHEMISTRY OF NITROGEN - A REVIEW. *Analyst* **109**, 549-568 (1984).
40. Bolleter, W.T., Bushman, C.J. & Tidwell, P.W. Spectrophotometric Determination of Ammonia as Indophenol. *Analytical Chemistry* **33**, 592-594 (1961).
41. Zhao, Y. et al. Ammonia Detection Methods in Photocatalytic and Electrocatalytic Experiments: How to Improve the Reliability of NH3 Production Rates? *Advanced Science* **6**, 1802109 (2019).
42. Yao, Y., Zhu, S., Wang, H., Li, H. & Shao, M. A Spectroscopic Study of Electrochemical Nitrogen and Nitrate Reduction on Rhodium Surfaces. *Angewandte Chemie International Edition* **n/a**.
43. Yao, Y., Zhu, S., Wang, H., Li, H. & Shao, M. A Spectroscopic Study on the Nitrogen Electrochemical Reduction Reaction on Gold and Platinum Surfaces. *Journal of the American Chemical Society* **140**, 1496-1501 (2018).

44. Buckley, K. & Ryder, A.G. Applications of Raman Spectroscopy in Biopharmaceutical Manufacturing: A Short Review. *Applied Spectroscopy* **71**, 1085-1116 (2017).

45. Gardiner, D., Hester, R. & Grossman, W. Ammonia in the liquid state and in solution: a Raman study. *Journal of Raman Spectroscopy* **1**, 87-95 (1973).

46. Simonelli, D. & Shultz, M.J. Temperature Dependence for the Relative Raman Cross Section of the Ammonia/Water Complex. *Journal of Molecular Spectroscopy* **205**, 221-226 (2001).

47. Sosa, C.P., Carpenter, J.E. & Novoa, J.J. in Chemical Applications of Density-Functional Theory, Vol. 629 131-141 (American Chemical Society, 1996).

48. Yeo, G.A. & Ford, T.A. The matrix isolation infrared spectrum of the water—ammonia complex. *Spectrochimica Acta Part A: Molecular Spectroscopy* **47**, 485-492 (1991).

49. Aggarwal, R.L., Farrar, L.W., Cecca, S.D. & Jeys, T.H. Raman spectra and cross sections of ammonia, chlorine, hydrogen sulfide, phosgene, and sulfur dioxide toxic gases in the fingerprint region 400-1400 cm−1. *AIP Advances* **6**, 025310 (2016).

50. Langseth, A. Feinstruktur von Ramanbanden - II. Das Ramanspektrum von Ammoniak in wässeriger Lösung. *Zeitschrift für Physik* **77**, 60-71 (1932).

51. Mysen, B.O. & Fogel, M.L. Nitrogen and hydrogen isotope compositions and solubility in silicate melts in equilibrium with reduced (N+H)-bearing fluids at high pressure and temperature: Effects of melt structure. *American Mineralogist* **95**, 987-999 (2010).

52. Zhao, Y. & Truhlar, D.G. The M06 suite of density functionals for main group thermochemistry, thermochemical kinetics, noncovalent interactions, excited states, and transition elements: two new functionals and systematic testing of four M06-class functionals and 12 other functionals. *Theoretical Chemistry Accounts* **120**, 215-241 (2008).

53. Blackie, E.J., Le Ru, E.C. & Etchegoin, P.G. Single-Molecule Surface-Enhanced Raman Spectroscopy of Nonresonant Molecules. *Journal of the American Chemical Society* **131**, 14466-14472 (2009).

54. Yampolsky, S. et al. Seeing a single molecule vibrate through time-resolved coherent anti-Stokes Raman scattering. *Nature Photonics* **8**, 650-656 (2014).

55. Sanchez, L.A., Lombardi, J.R. & Birke, R.L. Surface enhanced Raman scattering of ammonia. *Chemical Physics Letters* **108**, 45-50 (1984).

56. Farrell, M.E., Holthoff, E.L. & Pellegrino, P.M. Surface-Enhanced Raman Scattering Detection of Ammonium Nitrate Samples Fabricated Using Drop-on-Demand Inkjet Technology. *Applied Spectroscopy* **68**, 287-296 (2014).

57. Nazemi, M., Soule, L., Liu, M. & El-Sayed, M.A. Ambient Ammonia Electrosynthesis from Nitrogen and Water by Incorporating Palladium in Bimetallic Gold–Silver Nanocages. *Journal of The Electrochemical Society* **167**, 054511 (2020).

58. Nazemi, M. (Georgia Institute of Technology, 2020).

59. Schmidt, M.S., Hübner, J. & Boisen, A. Large Area Fabrication of Leaning Silicon Nanopillars for Surface Enhanced Raman Spectroscopy. *Advanced Materials* **24**, OP11-OP18 (2012).

60. Faiman, R. & Larsson, K. Assignment of the C□ H stretching vibrational frequencies in the Raman spectra of lipids. *Journal of Raman Spectroscopy* **4**, 387-394 (1976).

61. Yu, Y. et al. Complete Raman spectral assignment of methanol in the C–H stretching region. *The Journal of Physical Chemistry A* **117**, 4377-4384 (2013).

62. M. J. Frisch, G.W.T., H. B. Schlegel, G. E. Scuseria, M. A. Robb, J. R. Cheeseman, G. Scalmani, V. Barone, B. Mennucci, G. A. Petersson, H. Nakatsuji, M. Caricato, X. Li, H. P. Hratchian, A. F. Izmaylov, J. Bloino, G. Zheng, J. L. Sonnenberg, M. Hada, M. Ehara, K. Toyota, R. Fukuda, J. Hasegawa, M. Ishida, T. Nakajima, Y. Honda, O. Kitao, H. Nakai, T. Vreven, J. A. Montgomery, Jr., J. E. Peralta, F. Ogliaro, M. Bearpark, J. J. Heyd, E. Brothers, K. N. Kudin, V. N. Staroverov, T. Keith, R. Kobayashi, J. Normand, K. Raghavachari, A. Rendell, J. C. Burant, S. S. Iyengar, J. Tomasi, M. Cossi, N. Rega, J. M. Millam, M. Klene, J. E. Knox, J. B. Cross, V. Bakken, C. Adamo, J. Jaramillo, R. Gomperts, R. E. Stratmann, O. Yazyev, A. J. Austin, R. Cammi, C. Pomelli, J. W. Ochterski, R. L. Martin, K. Morokuma, V. G. Zakrzewski, G. A. Voth, P. Salvador, J. J. Dannenberg, S. Dapprich, A. D. Daniels, O. Farkas, J. B. Foresman, J. V. Ortiz, J. Cioslowski, and D. J. Fox, Edn. 01 (Gaussian, Inc., Wallington CT; 2010).

**Simple all-optical method for in situ detection of ultralow amount of ammonia**

Yuanchao Liu[a], Tristan F. Asset[a], Yechuan Chen[a], Eamonn Murphy[a], Eric O. Potma[b], Ivana Matanovic[c], Dmitry A. Fishman[b*], Plamen Atanassov[a]

[a]*Department of Chemical & Biomolecular Engineering, National Fuel Cell Research Center (NFCRC), University of California, Irvine, CA, 92697, USA*

[b]*Department of Chemistry, University of California, Irvine, California 92697, USA*

[c]*Department of Chemical and Biological Engineering, University of New Mexico, Albuquerque, New Mexico 87131, USA*

Supplementary information

* Corresponding author: dmitryf@uci.edu

| Formation | Spectral position | Intensity | Assignment |
|---|---|---|---|
| $NH_3$ | 3409.0 $cm^{-1}$ | 36.8 | NH asymmetric stretching |
| | 3309.8 $cm^{-1}$ | 155.8 | NH symmetric stretching |
| | 3309.8 $cm^{-1}$ | 155.8 | NH symmetric stretching |
| | 1660.4 $cm^{-1}$ | 1.25 | H-N-H bending |
| | 1660.0 $cm^{-1}$ | 1.25 | H-N-H bending |
| | 1033.1 $cm^{-1}$ | 0.57 | umbrella motion |
| | 3177.2 $cm^{-1}$ | | bending overtones |
| | 3190.7 $cm^{-1}$ | | bending overtones |
| $NH_4^+$ | 3380.0 $cm^{-1}$ | 16.8 | NH asymmetric stretching |
| | 3276.8 $cm^{-1}$ | 101.0 | NH symmetric stretching |
| | 1717.4 $cm^{-1}$ | 3.5 | H-N-H bending |
| | 1717.9 $cm^{-1}$ | 3.5 | H-N-H bending |
| | 1476.4 $cm^{-1}$ | 146.7 | NH deformations |
| | 1476.3 $cm^{-1}$ | 146.7 | NH deformations |
| | 1476.1 $cm^{-1}$ | 146.7 | NH deformations |
| | 3332.3 $cm^{-1}$ | | H-N-H bending overtones |
| | 3334.5 $cm^{-1}$ | | H-N-H bending overtones |
| | 2838.4 $cm^{-1}$ | | NH deformation overtones |
| | 2851.9 $cm^{-1}$ | | NH deformation overtones |
| | 2860.6 $cm^{-1}$ | | NH deformation overtones |
| $NH_3$-$NH_3$ | 3441.0 $cm^{-1}$ | 76.2 | NH asymmetric stretching |
| | 3438.0 $cm^{-1}$ | 7.8 | NH asymmetric stretching |
| | 3402.3 $cm^{-1}$ | 48.2 | NH asymmetric stretching |
| | 3369.5 $cm^{-1}$ | 15.6 | NH asymmetric stretching |
| | 3315.4 $cm^{-1}$ | 341.6 | NH symmetric stretching |
| | 1675.0 $cm^{-1}$ | 0.0 | H-N-H bending modes |
| | 1665.4 $cm^{-1}$ | 4.5 | H-N-H bending modes |
| | 1663.7 $cm^{-1}$ | 1.6 | H-N-H bending modes |
| | 1646.7 $cm^{-1}$ | 0.0 | H-N-H bending modes |
| | 1077.9 $cm^{-1}$ | 1.64 | umbrella motion |
| | 1047.7 $cm^{-1}$ | 0.0 | umbrella motion |
| | 440.9 $cm^{-1}$ | 0.3 | deformation modes |
| | 222.7 $cm^{-1}$ | 0.0 | deformation modes |
| | 146.2 $cm^{-1}$ | 0.2 | deformation modes |
| | 101.0 $cm^{-1}$ | 0.0 | deformation modes |
| | 82.9 $cm^{-1}$ | 15.7 | deformation modes |
| | 77.9 $cm^{-1}$ | 207.7 | deformation modes |
| | 3150.2 $cm^{-1}$ | | H-N-H bending overtones |
| | 3131.6 $cm^{-1}$ | | H-N-H bending overtones |
| | 3267.2 $cm^{-1}$ | | H-N-H bending overtones |
| $NH_3$-$H_2O$ | 3882.1 $cm^{-1}$ | 54.1 | OH stretching* |
| | 3545.6 $cm^{-1}$ | 175.0 | OH stretching (H-bonded) |
| | 3410.9 $cm^{-1}$ | 35.3 | NH asymmetric stretching |
| | 3389.2 $cm^{-1}$ | 33.0 | NH asymmetric stretching |
| | 3305.0 $cm^{-1}$ | 155.31 | NH symmetric stretching |
| | 1664.9 $cm^{-1}$ | 0.9 | HNH-HOH bending |
| | 1659.0 $cm^{-1}$ | 1.7 | H-N-H bending |
| | 1641.7 $cm^{-1}$ | 1.0 | HNH-HOH bending |
| | 1085.5 $cm^{-1}$ | 0.72 | umbrella mode |
| | 684.5 $cm^{-1}$ | 0.0 | HNH-HOH deformations |
| | 445.2 $cm^{-1}$ | 0.3 | HNH-HOH deformations |
| | 199.1 $cm^{-1}$ | 0.1 | HNH-HOH deformations |
| | 177.4 $cm^{-1}$ | 0.1 | HNH-HOH deformations |
| | 162.4 $cm^{-1}$ | 35.3 | HNH-HOH deformations |
| | 71.1 $cm^{-1}$ | 66.7 | HNH-HOH deformations |
| | 3000.0 $cm^{-1}$ | | bending overtones |
| | 3089.6 $cm^{-1}$ | | bending overtones |
| | 3166.8 $cm^{-1}$ | | bending overtones |
| Solvated $NH_3$ | 3755.5 $cm^{-1}$ | 66.0 | OH stretching * |
| | 3739.4 $cm^{-1}$ | 100.7 | OH stretching* |
| | 3700.0 $cm^{-1}$ | 89.3 | OH stretching* |
| | 3659.6 $cm^{-1}$ | 83.8 | OH stretching* |
| | 3624.0 $cm^{-1}$ | 83.8 | OH stretching* |
| | 3601.4 $cm^{-1}$ | 63.8 | OH stretching* |
| | 3507.0 $cm^{-1}$ | 55.0 | OH stretching* |
| | 3597.3 $cm^{-1}$ | 51.0 | OH stretching* |
| | 3562.0 $cm^{-1}$ | 74.1 | OH stretching |
| | 3501.0 $cm^{-1}$ | 24.6 | OH stretching |
| | 3490.9 $cm^{-1}$ | 86.9 | OH stretching |
| | 3496.8 $cm^{-1}$ | 34.0 | OH stretching |
| | 3525.9 $cm^{-1}$ | 69.8 | OH stretching |
| | 3470.0 $cm^{-1}$ | 16.3 | OH stretching |
| | 3489.5 $cm^{-1}$ | 77.9 | OH stretching |
| | 3421.4 $cm^{-1}$ | 112.4 | OH stretching |
| | 3488.9 $cm^{-1}$ | 77.8 | OH stretching |
| | 3434.2 $cm^{-1}$ | 37.7 | asymmetric NH stretching |
| | 3376.3 $cm^{-1}$ | 38.6 | asymmetric NH stretching |
| | 3299.6 $cm^{-1}$ | 74.5 | OH stretching |
| | 3263.3 $cm^{-1}$ | 48.3 | OH stretching |

| | | |
|---|---|---|
| 3249.7 $cm^{-1}$ | 145.8 | symmetric NH stretching |
| 3225.9 $cm^{-1}$ | 65.7 | OH stretching |
| 3173.5 $cm^{-1}$ | 145.5 | OH stretching |
| 3058.5 $cm^{-1}$ | 81 | OH stretching |
| 2971.4 $cm^{-1}$ | 119.8 | OH stretching |
| 2966.9 $cm^{-1}$ | 194.9 | OH stretching |

**Table ST1.** Full spectrum of ammonia related formations from DFT calculations. *OH that is not H-bonded.

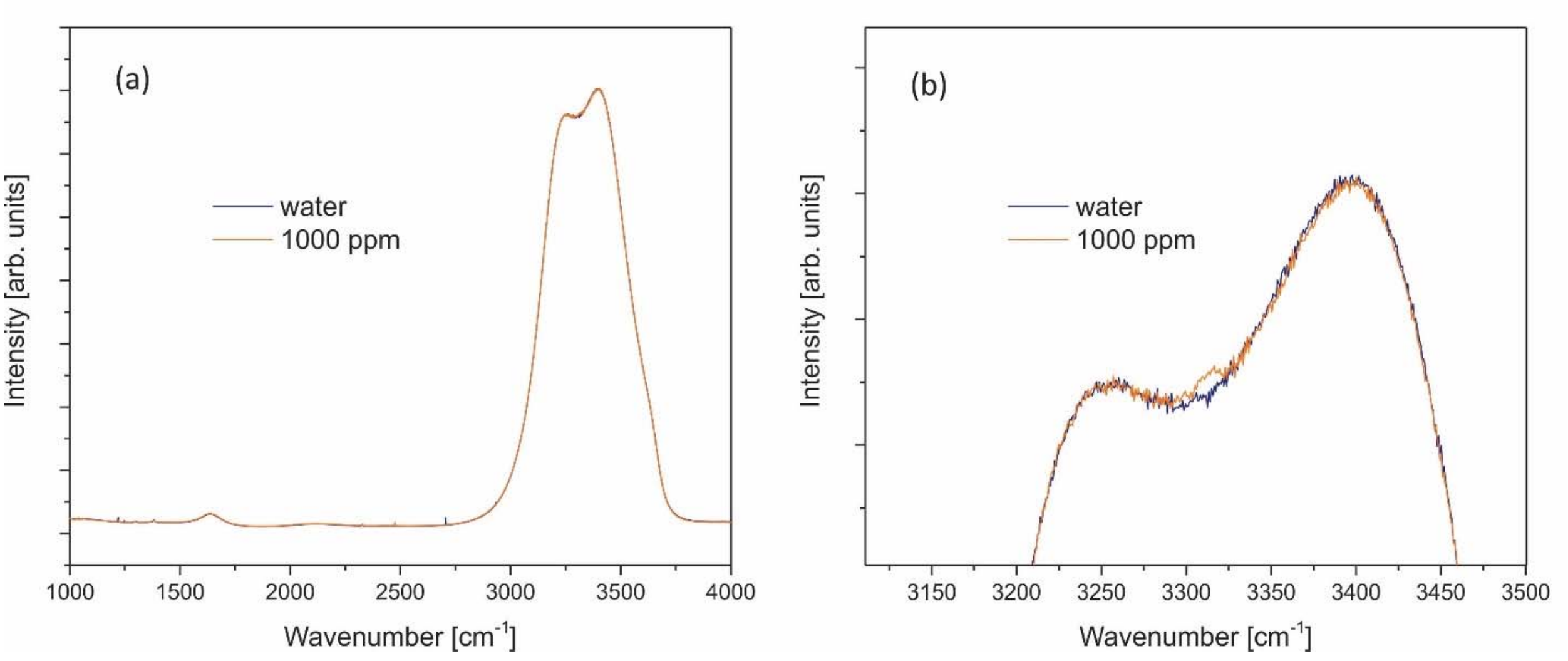

**Figure SF1.** (a) Spontaneous Raman spectra of water and 1000 ppm ammonia solution. (b) Zoomed part of spontaneous Raman spectra around $NH_3$ symmetric mode.

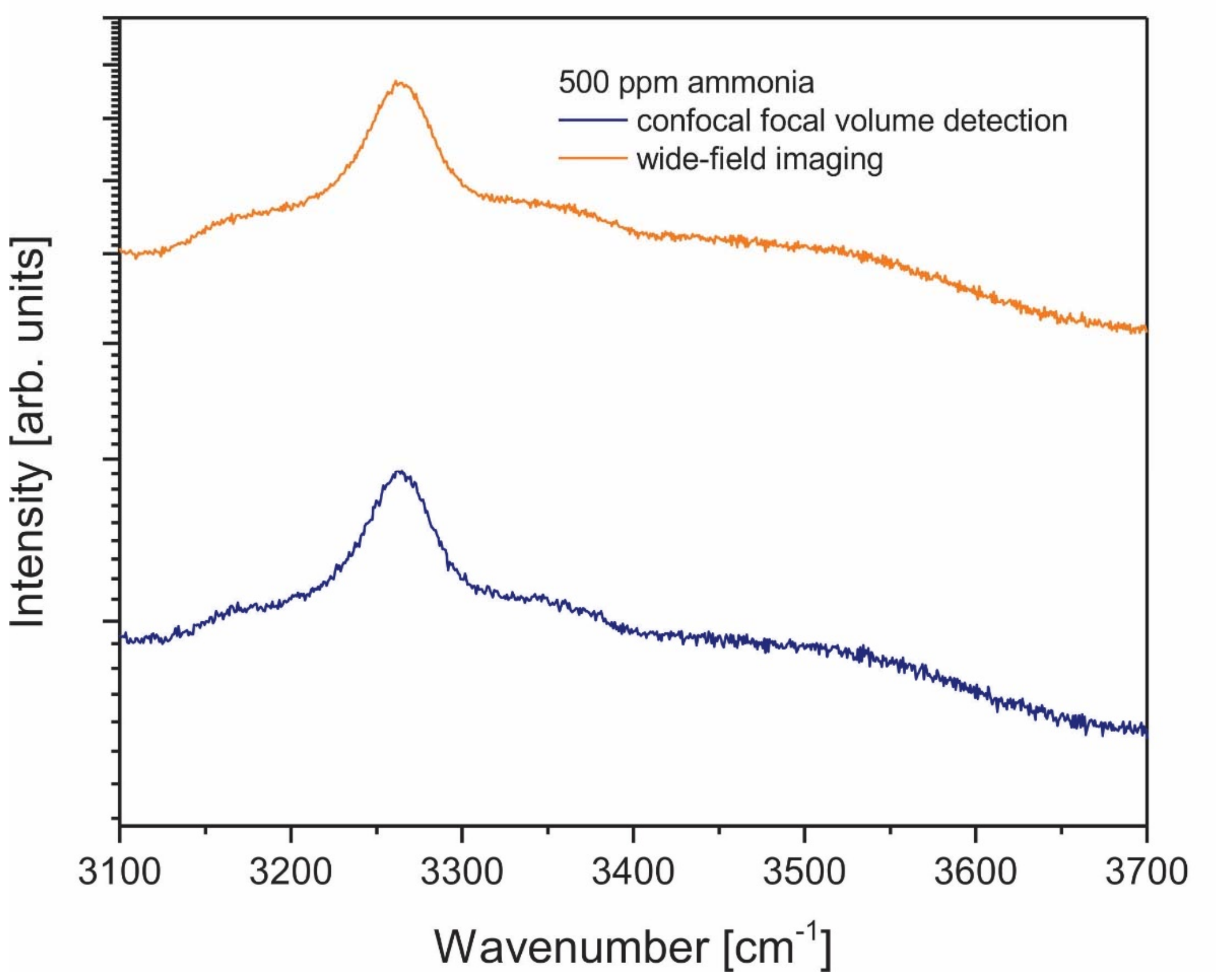

**Figure SF2.** Direct comparison of confocal imaging and wide-field detection approach.

| Formation | Parameter | Value |
|---|---|---|
| $NH_3$ | Spectral position | 3409.0 $cm^{-1}$ |
| | Amplitude | 36.8 |
| | Width | 200 $cm^{-1}$ |
| | Spectral position | 3309.8 $cm^{-1}$ |
| | Amplitude | 155.8 |
| | Width | 30 $cm^{-1}$ |
| $NH_3$-$H_2O$ | Spectral position | 3545.6 $cm^{-1}$ |
| | Amplitude | 175 |
| | Width | 200 $cm^{-1}$ |
| | Spectral position | 3410.9 $cm^{-1}$ |
| | Amplitude | 35.3 |
| | Width | 200 $cm^{-1}$ |
| | Spectral position | 3389.2 $cm^{-1}$ |
| | Amplitude | 33.0 |
| | Width | 200 $cm^{-1}$ |
| | Spectral position | 3305.0 $cm^{-1}$ |
| | Amplitude | 155.31 |
| | Width | 200 $cm^{-1}$ |

**Table ST2.** Parameters used for simple Gaussian visualization. No fitting has been done.

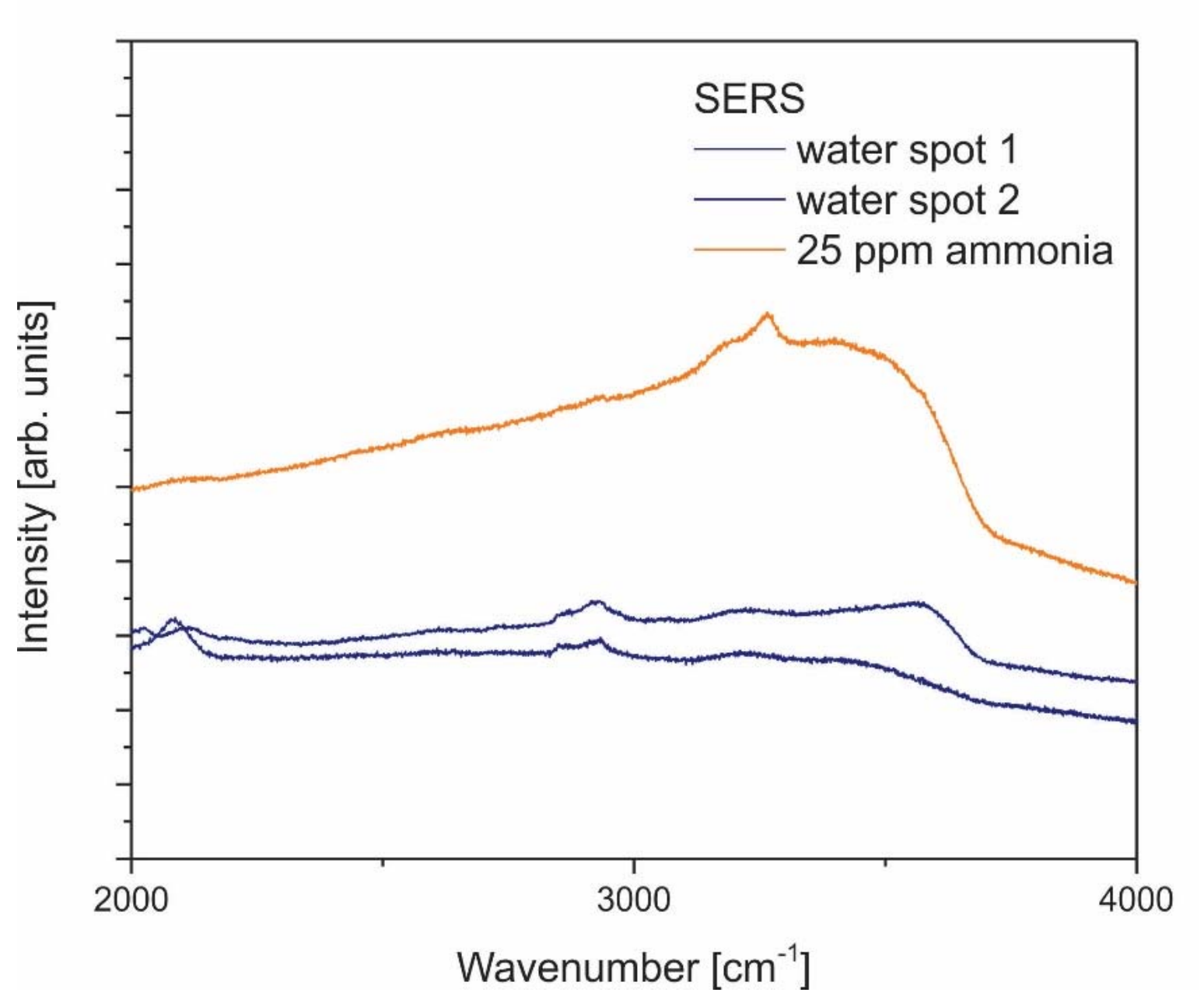

**Figure SF3.** SERS spectrum of 25 ammonia solution and water in different spots on substrate. Line at ~2900 $cm^{-1}$ has no concentration correlation similar to line at 3260 $cm^{-1}$.

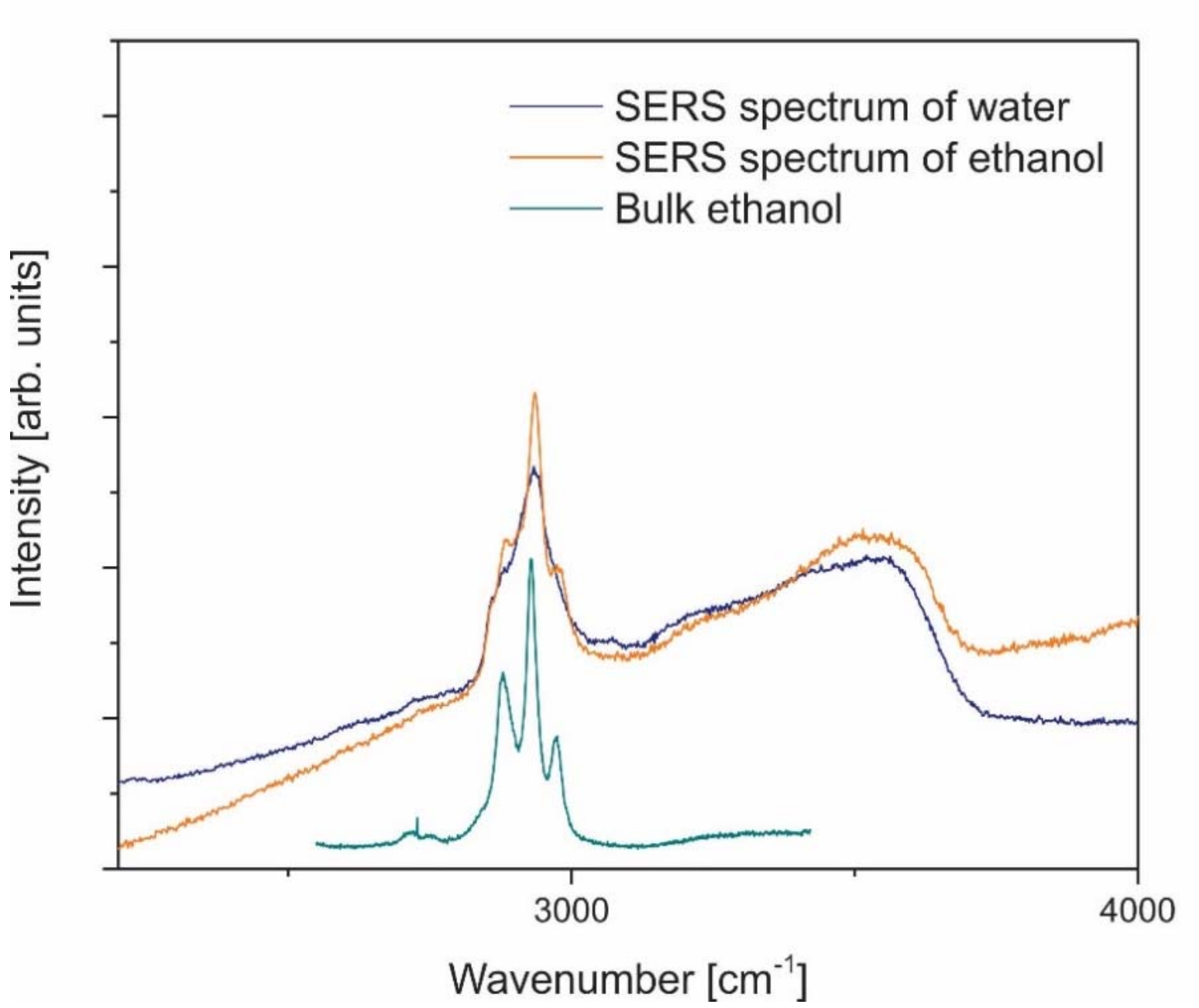

**Figure SF4.** Normalized SERS spectrum of water, ethanol overlaid with spontaneous Raman spectrum of ethanol. Line position and spectral structure indicates contamination of SERS substrate with organic compound sporadically distributed on substrate. SERS spectra clearly shows broadened lines with respect to spontaneous Raman signals, which is expected for molecule-metal interaction.

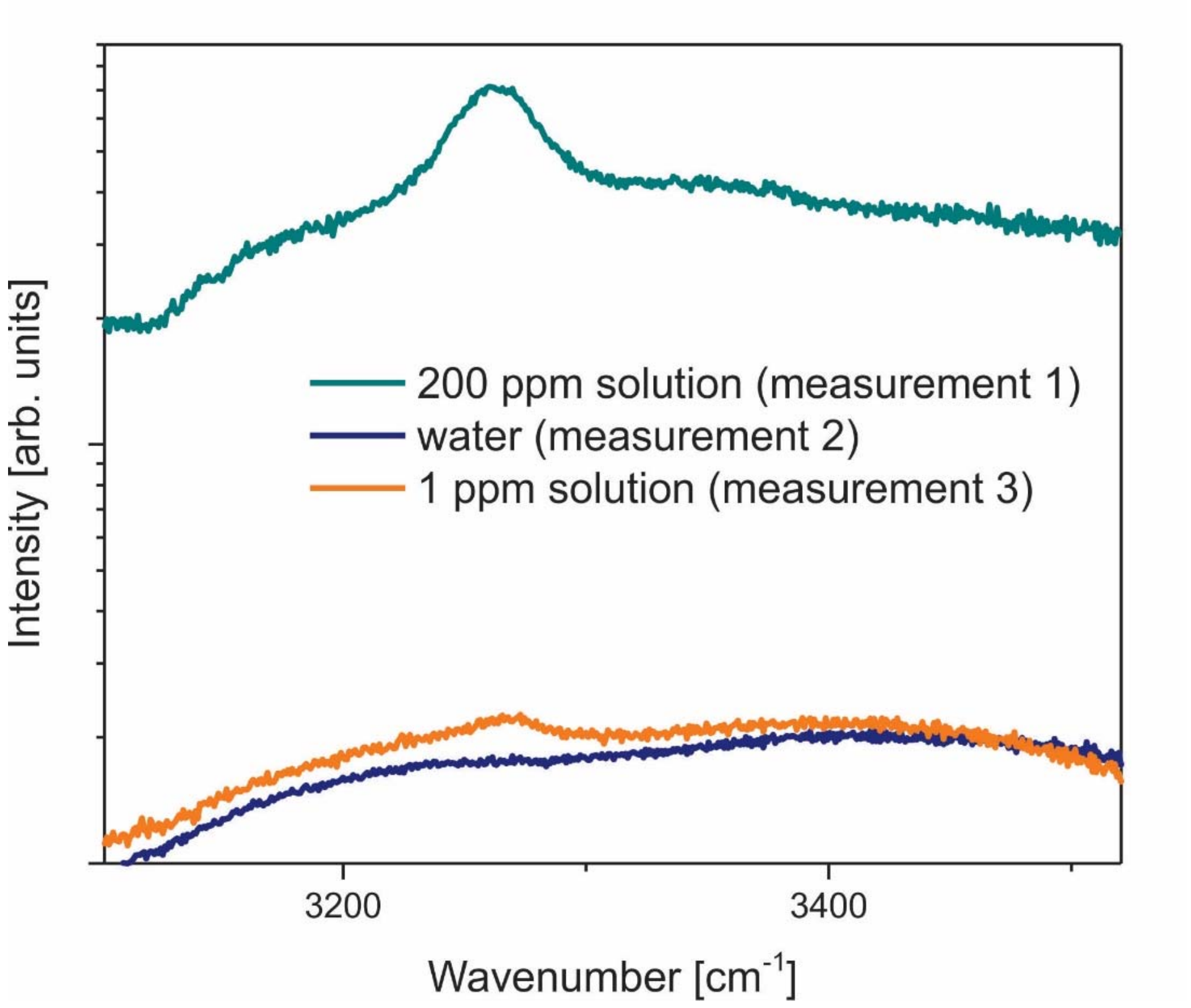

**Figure SF5.** SERS on Ag plasmonic structures allows detection of both increase and decrease of the amount the ammonia.

| p.p.b. | p.p.m. | mg/L | mmol/L | Number of molecules |
|---|---|---|---|---|
| 50 | 0.05 | 2.936 | 5*10-6 | 3106 |
| 100 | 0.1 | 5.872 | 1*10-5 | 6212 |
| 500 | 0.5 | 29.358 | 5*10-5 | 31062 |
| 1000 | 1 | 58.717 | 1*10-4 | 62124 |
| 2000 | 2 | 117.433 | 2*10-4 | 124249 |

**Table ST3.** Conversion table. Simple approximation of amount of molecule in confocal volume.